\documentclass[prx,aps,floatfix,longbibliography,superscriptaddress,twocolumn,final,notitlepage]{revtex4-2}
\usepackage{float}
\usepackage{graphicx}
\usepackage{subcaption}
\usepackage{cleveref}
\begin{document}

\title[
    Raphtory: The temporal graph engine for Rust and Python
    ]{
    Raphtory: The temporal graph engine for Rust and Python
}

\author{Ben Steer}
\email{ben.steer@pometry.com}
\affiliation{Pometry, United Kingdom}

\author{Naomi Arnold}
\email{naomi.arnold@nulondon.ac.uk}
\affiliation{Network Science Institute, Northeastern University London, United Kingdom}

\author{Cheick Tidiane Ba}
\affiliation{University of Milan, Italy}
\affiliation{School of Electronic Engineering and Computer Science, Queen Mary University of London, United Kingdom}

\author{Renaud Lambiotte}
\affiliation{Pometry, United Kingdom}
\affiliation{Mathematical Institute, University of Oxford, United Kingdom}
\affiliation{The Alan Turing Institute, United Kingdom}

\author{Haaroon Yousaf}
\affiliation{Pometry, United Kingdom}

\author{Lucas Jeub}
\affiliation{Pometry, United Kingdom}

\author{Fabian Murariu}
\affiliation{32 Bytes Software, United Kingdom}

\author{Shivam Kapoor}
\affiliation{Pometry, United Kingdom}

\author{Pedro Rico}
\affiliation{Pometry, United Kingdom}

\author{Rachel Chan}
\affiliation{Pometry, United Kingdom}

\author{Louis Chan}
\affiliation{Pometry, United Kingdom}

\author{James Alford}
\affiliation{Pometry, United Kingdom}

\author{Richard G. Clegg}
\affiliation{School of Electronic Engineering and Computer Science, Queen Mary University of London, United Kingdom}

\author{Felix Cuadrado}
\affiliation{Universidad Polit\'ecnica de Madrid, Spain}
\affiliation{School of Electronic Engineering and Computer Science, Queen Mary University of London, United Kingdom}

\author{Matthew Russell Barnes}
\affiliation{School of Electronic Engineering and Computer Science, Queen Mary University of London, United Kingdom}

\author{Peijie Zhong}
\affiliation{School of Electronic Engineering and Computer Science, Queen Mary University of London, United Kingdom}

\author{John N. Pougu\'e Biyong}
\affiliation{Mathematical Institute, University of Oxford, United Kingdom}

\author{Alhamza Alnaimi}
\affiliation{Pometry, United Kingdom}

\maketitle

\section*{Summary}

Raphtory is a platform for building and analysing temporal networks. The library includes methods for creating networks from a variety of data sources; algorithms to explore their structure and evolution; and an extensible GraphQL server for deployment of applications built on top. Raphtory's core engine is built in Rust, for efficiency, with Python interfaces, for ease of use. Raphtory is developed by network scientists, with a background in Physics, Applied Mathematics, Engineering and Computer Science, for use across academia and industry. 
\section*{Statement of need}
Networks are at the core of data science solutions in a range of domains, including computer science, computational social science, and the life sciences~\cite{newman2018networks}. Networks are a powerful language focusing on the connectivity of systems, and offer a rich toolbox to extract greater understanding from data. Several network analysis tools exist, including NetworkX~\cite{hagberg2008exploring}, graph-tool~\cite{peixoto2014graph}  and igraph~\cite{csardi2006igraph}, and are freely accessible to scientists, practitioners and data miners. 

However, with abundant cheap storage and tools for logging every event which occurs in an ecosystem, datasets have become increasingly rich, combining different types of information that cannot be incorporated in a standard network model~\cite{lambiotte2019networks}. In particular, the temporal nature of many complex systems has led to the emergence of the field of temporal networks, with its own models and algorithms~\cite{holme2012temporal,masuda2016guide}.

Unfortunately, despite active academic research in the last decade, no efficient, generalised and production-ready system has been developed to explore the temporal dimension of networks. To support practitioners who wish to exploit both the structure and dynamics of their data, we have developed Raphtory.
\section*{Related software}
Besides the aforementioned packages, few open access tools have been developed for the mining of temporal networks, with the existing solutions focusing on specific sub-problems within the space. Those which have attempted to generalise to all temporal network analysis are either actively under development, but too preliminary to use in production, or have been abandoned due to lack of funding or changing research goals. 

As examples of these three categories: Pathpy is a Python package for the analysis of time series data on networks, but focuses on extracting and analysing time-respecting paths~\cite{hackl2021analysis}. Similarly DyNetX~\cite{DyNetX}, a pure python library relying on networkX, focuses on temporal slicing and the computation of time-respecting paths. The recently released Reticula offers a range of methods developed in C++ with a Python interface~\cite{badie2023reticula}. Phasik~\cite{lucas2023inferring}, written in Python, focuses on inferring phases from temporal network data. EvolvingGraphs.jl~\cite{zhang2015dynamic}, RecallGraph~\cite{RecallGraph} and Chronograph~\cite{Chronograph} all saw significant work before development was halted indefinitely.

Raphtory is a valuable addition to this ecosystem for the following reasons. Originally developed in Scala~\cite{steer2020raphtory}, its current core is entirely written in Rust. This is to ensure fast and memory-efficient computation that a pure python implementation could not achieve, and to handle the sheer volume of temporal network data, which often dwarfs that of an equivalent static network.

The library provides an expressive Python interface for interoperability with other data science tools, as well as simpler and more maintainable code. In addition, the library is built with a focus on scalability, as it relies on efficient data structures that can be used to extract different views of large temporal graphs. This avoids the creation of multiple graph objects that is not feasible with large datasets. The use of these new features is supported by well-documented APIs and tutorials, guiding the user from data loading through to analysis.

\section*{Overview}

\begin{figure*}[htbp]
\centering
    \begin{subfigure}[b]{\textwidth}
        \centering
        \includegraphics[width=0.45\linewidth]{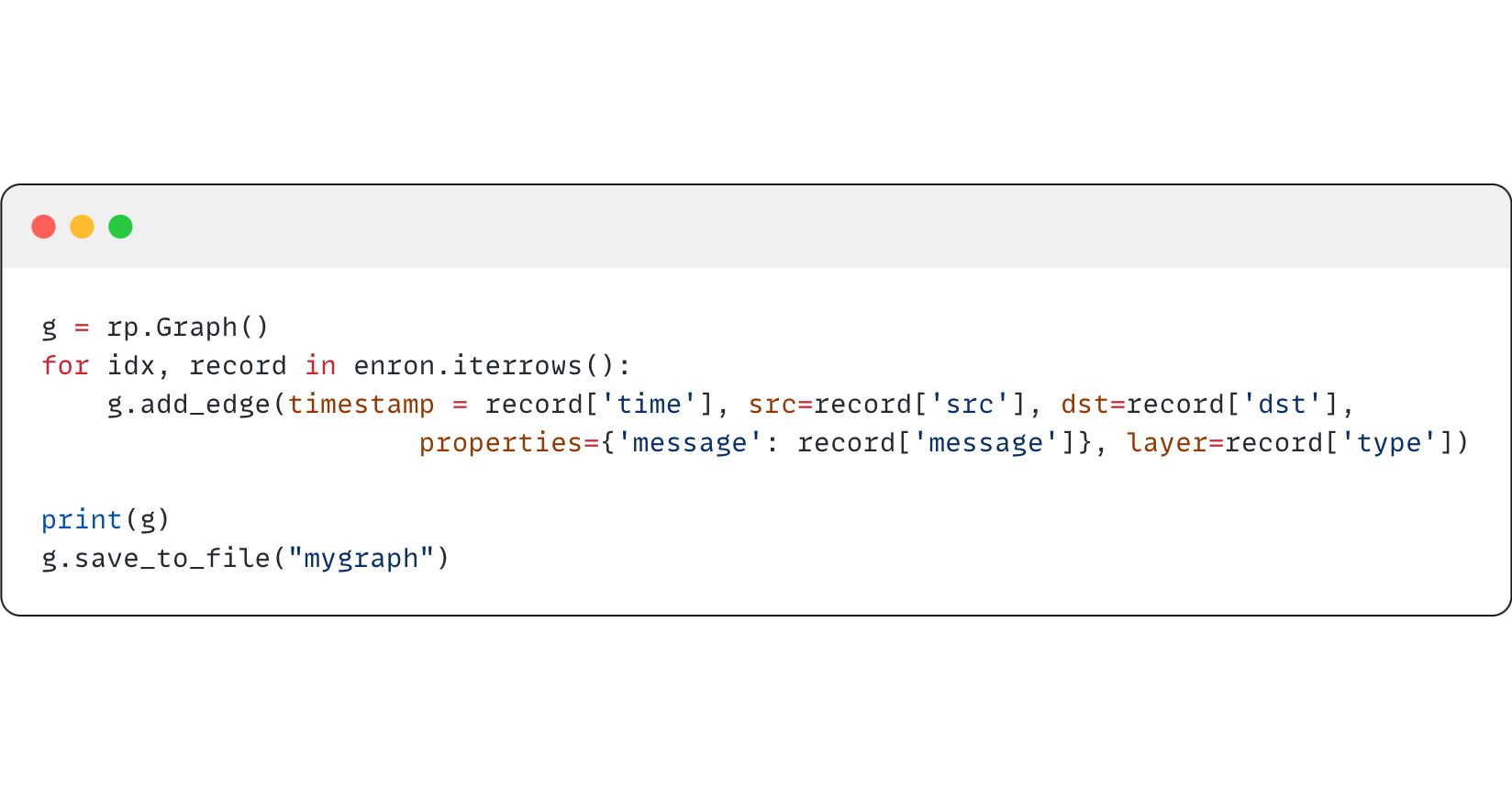}
        \hfill
        \includegraphics[width=0.45\linewidth]{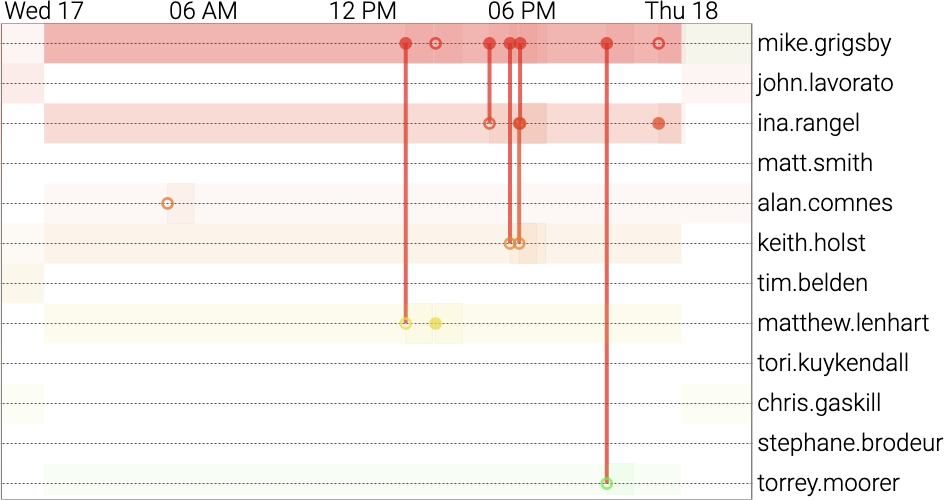}
        \caption{In a temporal network, edges are dynamical entities connecting pairs of nodes. Left is code for creating an example temporal network in Raphtory from a dataset of emails, right is a visual schematic containing this information.}
        \label{fig:ingest}
    \end{subfigure} \\
    \begin{subfigure}[b]{\textwidth}
        \centering
        \includegraphics[width=0.45\linewidth]{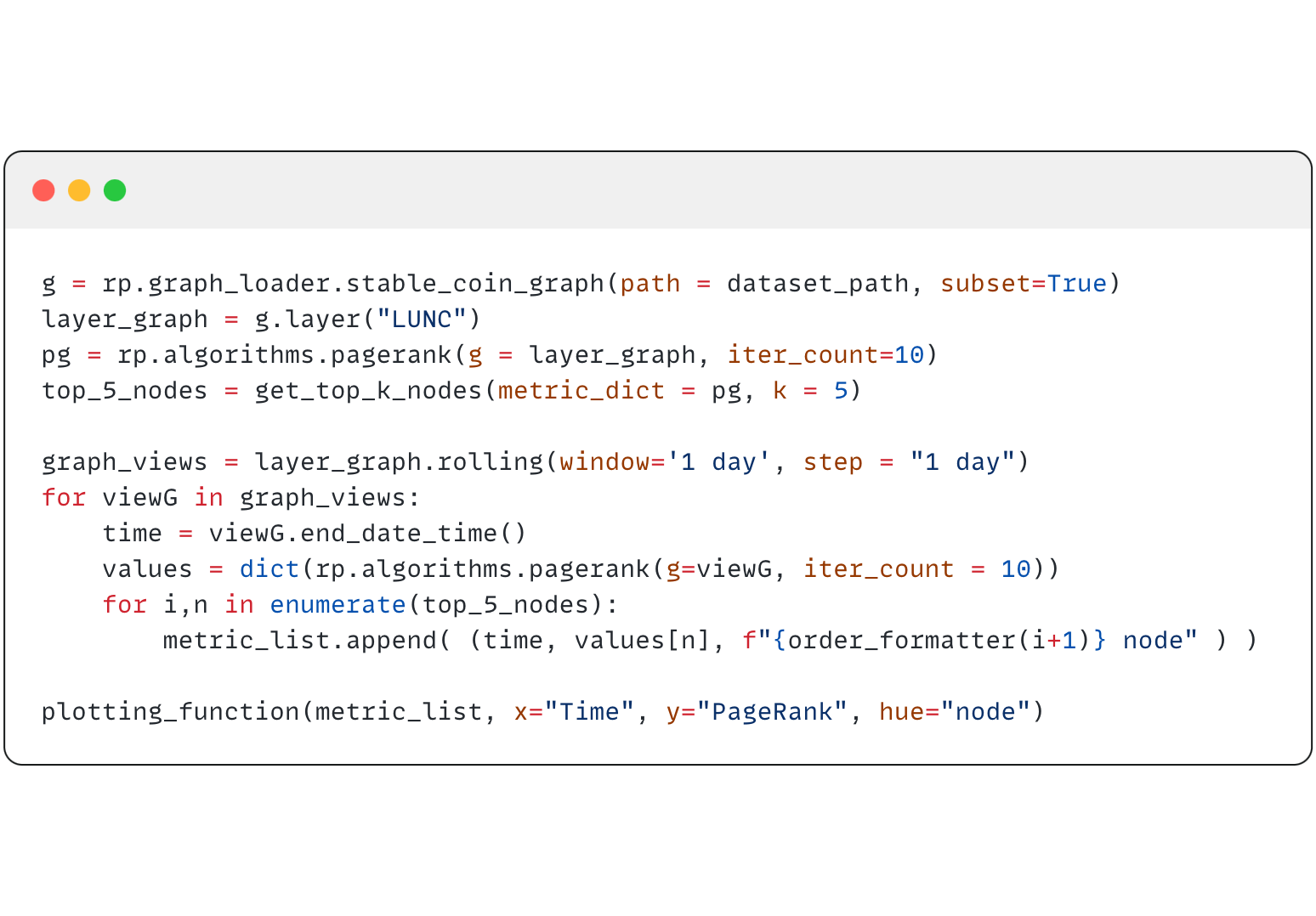}
        \hfill
        \includegraphics[width=0.45\linewidth]{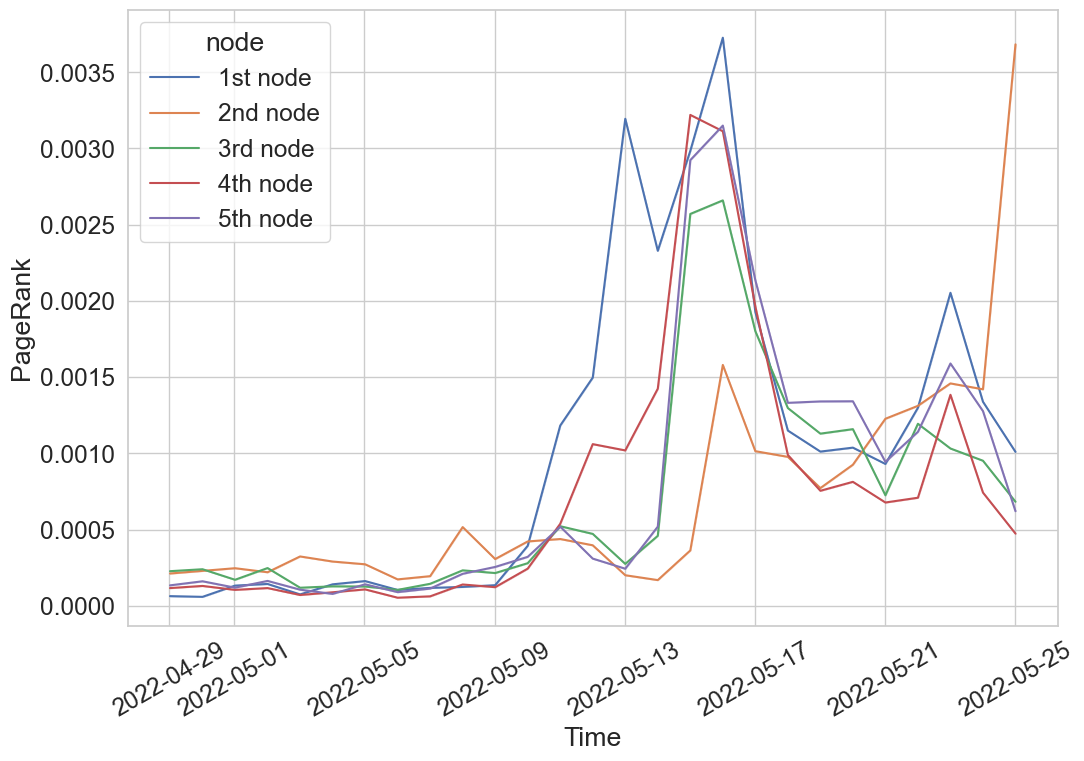}
        \caption{Generation of a sequence of graph views at a given time resolution and on selected layers, to run standard network algorithms, here Pagerank. Left: the all-time top five nodes by PageRank are selected, then their PageRank over a set of monthly sliding windows is calculated. Right: these values are plotted over time in a dataset of Ethereum transactions. }
        \label{fig:windows}
    \end{subfigure} \\
    \begin{subfigure}[b]{\textwidth}
        \centering
        \includegraphics[width=0.45\linewidth]{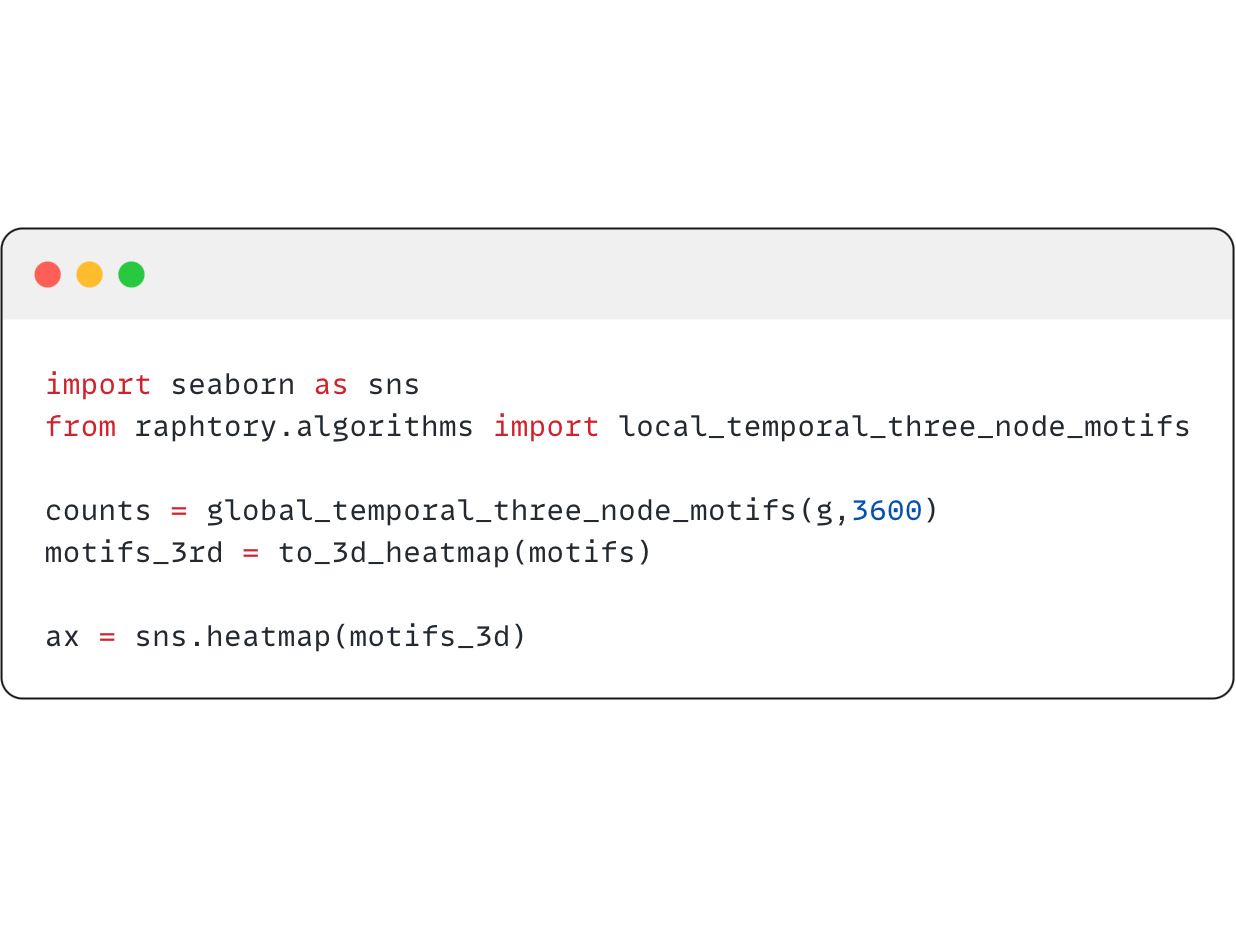}
        \includegraphics[width=0.45\linewidth]{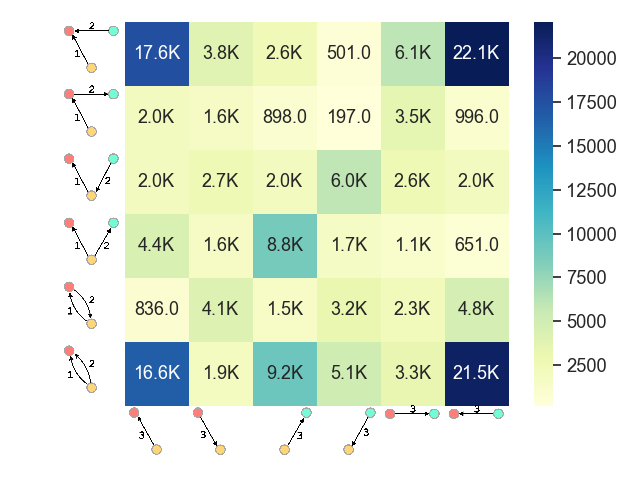}
        \caption{Raphtory  offers rapid implementations of algorithms specifically designed for temporal networks, here finding significant temporal motifs~\cite{paranjape2017motifs}. Left shows the code required for extracting all three-edge up-to-three node motifs from a dataset of StackExchange interactions that complete within 1 hour (3600 seconds), right is a visualisation of these motif counts (the row is the first two edges of the motif, the column the third edge).}
        \label{fig:motifs}
    \end{subfigure} \\
    \caption{Samples from different domains showcasing Raphtory's feature set.} \label{fig:sx-example}
\end{figure*}

The core Raphtory model consists of a base temporal graph which maintains a chronological log of all changes to its structure and property values over time. A graph can be created using simple functions for adding/removing vertices and edges at different time points, as well as updating their properties. Alternatively, a graph can be generated through in-built loaders for common data sources/formats (\cref{fig:ingest}). 

Once a graph has been created, a user may generate 'graph views' which set some structural or temporal constraints through which the underlying graph may be observed. Graph views can be generated programmatically over a desired time range (windows), over sets of nodes which pass some user-defined criteria (subgraphs), or over a subset of layers if the graph is multilayered. Additionally, the views can leverage event durations and support various semantics for deletions.
To reduce memory footprint, graph views are only materialised upon access. This allows a user to maintain thousands of different perspectives of their graph simultaneously, which can be explored and compared through the application of graph algorithms and metrics (\cref{fig:windows}). Furthermore, Raphtory provides extensions for null model generation and exporting of views to other graph libraries such as NetworkX.   

Raphtory includes fast and scalable implementations of algorithms for temporal network mining such as temporal motifs (\cref{fig:motifs}) and temporal reachability. In addition, it exposes its internal API for implementing algorithms in Rust, and surfacing them in Python.

Finally, Raphtory is built with a focus on ease of use and can be installed using standard Python and Rust package managers. Once installed it can be integrated within an analysis pipeline or run standalone as a GraphQL service.
\section*{Projects using Raphtory}
Raphtory has proved an invaluable resource in industrial and academic   projects, for instance to characterise the time evolution of the fringe social network Gab ~\cite{arnold2021moving}, transactions of users of a dark web marketplace Alphabay using temporal motifs~\cite{paranjape2017motifs} or anomalous patterns of activity in NFT trades \cite{yousaf2023non}. The library has recently been significantly rewritten, and we expect that with its new functionalities, efficiency and ease of use, it will become an essential part of the network scince community.
\section*{Funding}
R.L. acknowledges support from the EPSRC Grants
EP/V013068/1 and EP/V03474X/1. This work is part of the project ``Raphtory: a practical system for the analysis of dynamic graphs'' funded by the Alan Turing Institute.

\bibliographystyle{ScienceAdvances}
\bibliography{joss-raphtory}

\begin{thebibliography}{10}

\bibitem{newman2018networks}
M.~Newman, {\it Networks\/} (Oxford University Press, 2018).

\bibitem{hagberg2008exploring}
A.~Hagberg, P.~Swart, D.~S~Chult, Exploring network structure, dynamics, and
  function using networkx, {\it Tech. rep.\/}, Los Alamos National Lab.(LANL),
  Los Alamos, NM (United States) (2008).

\bibitem{peixoto2014graph}
T.~P. Peixoto, The graph-tool python library.
\newblock {\it figshare\/}  (2014).

\bibitem{csardi2006igraph}
G.~Csardi, T.~Nepusz, {\it et~al.\/}, The igraph software package for complex
  network research.
\newblock {\it InterJournal, complex systems\/} {\bf 1695}, 1--9 (2006).

\bibitem{lambiotte2019networks}
R.~Lambiotte, M.~Rosvall, I.~Scholtes, From networks to optimal higher-order
  models of complex systems.
\newblock {\it Nature physics\/} {\bf 15}, 313--320 (2019).

\bibitem{holme2012temporal}
P.~Holme, J.~Saram{\"a}ki, Temporal networks.
\newblock {\it Physics reports\/} {\bf 519}, 97--125 (2012).

\bibitem{masuda2016guide}
N.~Masuda, R.~Lambiotte, {\it A guide to temporal networks\/} (World
  Scientific, 2016).

\bibitem{hackl2021analysis}
J.~Hackl, I.~Scholtes, L.~V. Petrovi{\'c}, V.~Perri, L.~Verginer, C.~Gote, {\it
  Companion Proceedings of the Web Conference 2021\/} (2021), pp. 530--532.

\bibitem{DyNetX}
G.~Rossetti, E.~ter Hoeven, U.~Norman, D.~Jorquera, H.~Dormán, M.~Dorner,
  Giuliorossetti/dynetx: v0.3.2 (2023).

\bibitem{badie2023reticula}
A.~Badie-Modiri, M.~Kivel{\"a}, Reticula: A temporal network and hypergraph
  analysis software package.
\newblock {\it SoftwareX\/} {\bf 21}, 101301 (2023).

\bibitem{lucas2023inferring}
M.~Lucas, A.~Morris, A.~Townsend-Teague, L.~Tichit, B.~Habermann, A.~Barrat,
  Inferring cell cycle phases from a partially temporal network of protein
  interactions.
\newblock {\it Cell Reports Methods\/} {\bf 3} (2023).

\bibitem{zhang2015dynamic}
W.~Zhang, Dynamic network analysis in {Julia}  (2015).

\bibitem{RecallGraph}
A.~Mukhopadhyay, Recallgraph, \url{https://github.com/RecallGraph/RecallGraph}
  (Accessed 19-06-2023).

\bibitem{Chronograph}
B.~Erb, D.~Mei\ss{}ner, J.~Pietron, F.~Kargl, {\it Proceedings of the 11th ACM
  International Conference on Distributed and Event-Based Systems\/} (2017).

\bibitem{steer2020raphtory}
B.~Steer, F.~Cuadrado, R.~Clegg, Raphtory: Streaming analysis of distributed
  temporal graphs.
\newblock {\it Future Generation Computer Systems\/} {\bf 102}, 453--464
  (2020).

\bibitem{paranjape2017motifs}
A.~Paranjape, A.~R. Benson, J.~Leskovec, {\it Proceedings of the tenth ACM
  international conference on web search and data mining\/} (2017), pp.
  601--610.

\bibitem{arnold2021moving}
N.~A. Arnold, B.~Steer, I.~Hafnaoui, H.~A. Parada~G, R.~J. Mondrag{\'o}n,
  F.~Cuadrado, R.~G. Clegg, Moving with the times: Investigating the alt-right
  network {Gab} with temporal interaction graphs.
\newblock {\it Proceedings of the ACM on Human-Computer Interaction\/}  (2021).

\bibitem{yousaf2023non}
H.~Yousaf, N.~A. Arnold, R.~Lambiotte, T.~LaRock, R.~G. Clegg, P.~Zhong,
  A.~Alnaimi, B.~Steer, Non-markovian paths and cycles in {NFT} trades.
\newblock {\it arXiv preprint arXiv:2303.11181\/}  (2023).

\end{thebibliography}

\end{document}